\documentclass[onecollarge]{svjour2}       
\smartqed  
\usepackage{graphicx}
\usepackage{amsmath}

\journalname{Journal of Statistical Physics}

\begin{document}

\title{{Statistical Mechanics of the Minimum Dominating Set 
    Problem}\thanks{Research partially supported by the National Basic 
    Research Program of China (grant number 2013CB932804) and by the National
    Natural Science Foundation of China (grand numbers 11121403 and 11225526).}}

\titlerunning{Minimum dominating set}

\author{Jin-Hua Zhao \and Yusupjan Habibulla \and Hai-Jun Zhou}

\institute{J.-H. Zhao \and Y. Habibulla \and H.-J. Zhou \at
  State Key Laboratory of Theoretical Physics,
  Institute of Theoretical Physics,
  Chinese Academy of Sciences, \\
  Zhong-Guan-Cun East Road 55, Beijing 100190, China.
  (\email{zhouhj@itp.ac.cn})
}

\date{February 05, 2015}

\maketitle

\begin{abstract}
  The minimum dominating set (MDS) problem has wide applications in network
  science and related fields. It aims at constructing a node set
  of smallest size such that any node of the network is either in this
  set or is adjacent to at least one node of this set. Although this 
  optimization problem is 
  generally very difficult, we show it can be exactly 
  solved by a generalized leaf-removal (GLR) process if the network contains
  no core. We present a percolation theory to describe the GLR process on 
  random networks, and solve a spin glass model by mean field method
  to estimate the MDS size. 
  We also implement a message-passing algorithm and
  a local heuristic algorithm that combines GLR with greedy 
  node-removal 
  to obtain near-optimal solutions for single random networks.
  Our algorithms also perform well on real-world network instances.
  \keywords{dominating set \and spin glass \and core percolation \and
    leaf removal \and network coarse-graining \and belief propagation}
\end{abstract}

\section{Introduction}

\begin{figure}
  \begin{center}
    \includegraphics[width=0.8\textwidth]{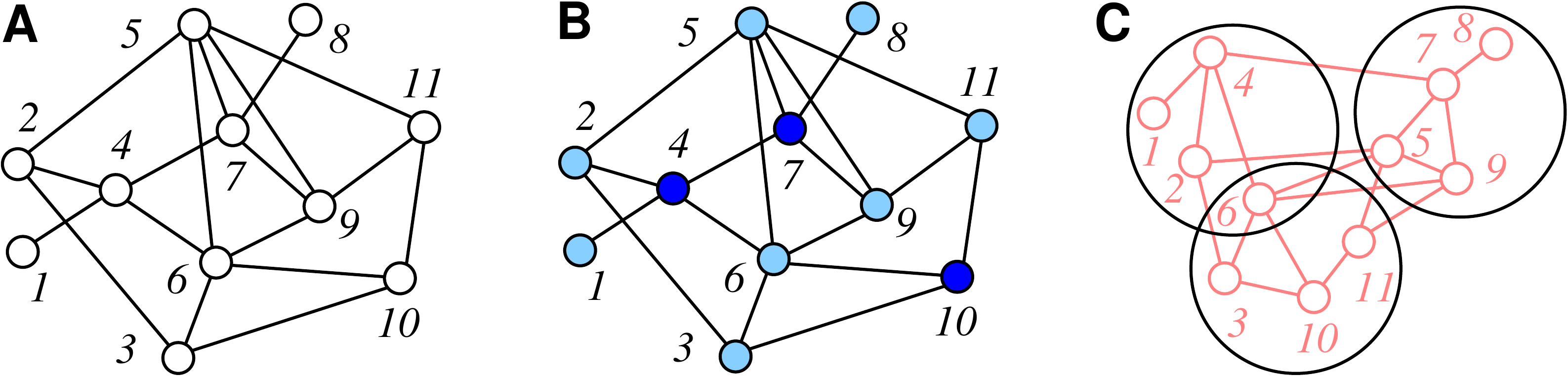}
  \end{center}
  \caption{
    \label{fig:MDSexample}
    An example of minimum dominating set. (A) a small network with
    $N=11$ nodes and $M=18$ links. (B) blue (dark gray) 
    indicates a node being occupied, while cyan (light gray)
    indicates a node being empty but observed.
    The three occupied nodes form a MDS $\Gamma_0=\{4,7,10\}$ 
    for this network.
    (C) a coarse-grained representation of the network based on the
    MDS of (B).
  }
\end{figure}

The minimum dominating set (MDS) problem 
\cite{Haynes-Hedetniemi-Slater-1998} has fundamental importance in 
network science. For example,  to ensure the proper functioning of a 
complex networked system such as a nation-wide power grid, it is often 
necessary to monitor the system's
microscopic dynamics in real-time by placing sensors on the nodes.
A sensor may have the capability of observing the instantaneous states of 
the residing node and all its adjacent nodes in the network
\cite{Yang-Wang-Motter-2012}, so they may not need to occupy all the nodes.
We then have the MDS problem: How to place sensors on as few
nodes as possible to minimize costs but still ensure that each node
is either occupied or adjacent to at least one occupied node?
As an example we show in Fig.~\ref{fig:MDSexample}B
a minimum dominating set (containing only three nodes)
of a small network.
A more stringent constraint, which is adopted in lattice glass
models \cite{Biroli-Mezard-2002}, is to require an empty node $i$ to be 
surrounded by at least $l_i$ occupied nodes, with $l_i$ being 
node-dependent. The MDS problem corresponds to
the case of $l_i \equiv 1$, while the other limiting case of 
$l_i = d_i$ is just the vertex cover (or independent 
set) problem \cite{Hartmann-Weigt-2003,Zhao-Zhou-2014}, where $d_i$ is
node $i$'s degree (i.e., number of adjacent nodes).

The MDS problem has wide practical applications, such as monitoring 
large-scale power  grids and other transportation systems 
\cite{Yang-Wang-Motter-2012}, 
controlling the spreading of infectious diseases and other network
dynamical processes 
\cite{Echenique-etal-2005,Takaguchi-Hasegawa-Yoshida-2014,Liu-Slotine-Barabasi-2011,Wuchty-2014},
efficient routing in wireless networks \cite{Wu-Li-2001}, and
network public goods games (e.g., resource allocation)
\cite{Bramoulle-Kranton-2007}.  Another application is to build
a coarse-grained representation for a complex network starting from
a MDS.  Such an idea has already been applied to
multi-document summarization in the field of information
extraction \cite{Shen-Li-2010}.
Each node $i$ of the MDS can be regarded as a representative node for
a local domain of the network. We can take the subnetwork induced by node $i$
and all its adjacent nodes (except those in the MDS) 
as  a coarse-grained node, and set up an coarse-grained 
link  between two coarse-grained nodes if the two
corresponding subnetworks share at least one
node or are connected by at least one link in
the original network (see Fig.~\ref{fig:MDSexample}C for an example).
If such a coarse-graining process is iterated we will then obtain a
hierarchical representation for the original network,
which may be very useful for understanding the organization
of a complex system and for searching and information transmission
in such a system.

Exactly solving the MDS problem, however, 
is extremely difficult in general, 
since it is a nondeterministic 
polynomial-complete (NP-complete) optimization problem 
\cite{Haynes-Hedetniemi-Slater-1998}. Even the task of approximately
solving the MDS problem is very hard. For a general network of $N$ nodes, 
so far the best polynomial algorithms can only guarantee to get dominating sets
with sizes not exceeding $\ln N$ times of the minimum size  
\cite{Lund-Yannakakis-1994,Raz-Safra-1997}.
Many local-search algorithms have been proposed to solve the MDS problem
heuristically (see review \cite{Haynes-Hedetniemi-Slater-1998} and
\cite{Echenique-etal-2005,Yang-Wang-Motter-2012,Hedar-Ismail-2012,Takaguchi-Hasegawa-Yoshida-2014,Molnar-etal-2013,Wuchty-2014}),
but theoretical results on the MDS sizes of random network ensembles
are still very rare.

In this work we bring several new theoretical and algorithmic contributions.
We show in Sec.~\ref{sec:core} that a 
generalized leaf-removal (GLR) process may cause a
core percolation transition, and  propose a quantitative theory to describe
this percolation.  If the network contains no core, GLR reaches
an exact MDS; if an extensive core exists, we combine GLR with a 
local greedy process in Sec.~\ref{sec:hybrid}
to get an upper bound to the MDS size. 
We then introduce a spin glass model in Sec.~\ref{sec:glass}
and estimate the MDS size by a
replica-symmetric (RS) mean field theory, and implement a message-passing
algorithm in Sec.~\ref{sec:bpd}
to get near-optimal dominating sets for single random
network instances.
Our algorithms also perform well on real-world network 
instances. This work shall be useful both for network scientists
who are interested in applying the MDS concept to practical problems,
and  for applied mathematicians who seek better theoretical understanding
on the random MDS problem.

\section{Generalized Leaf-Removal and Core Percolation}
\label{sec:core}

Consider a simple 
network $W$ formed by $N$ nodes and $M$ undirected links, each link
connecting between two different nodes.
Each node with index $i\in \{1, 2, \ldots, N\}$ is either empty (indicated
by the occupation state $c_i=0$) or occupied by sensors ($c_i=1$).
A node $i$ is regarded as observed if it is occupied or it is 
empty but adjacent to one or more occupied nodes, otherwise it is regarded as
unobserved. 
We need to occupy a set $\Gamma$ of nodes to make all the $N$
nodes be observed, and the objective is to make the 
dominating set $\Gamma$ as small as possible, i.e., to construct a
minimum dominating set. It is easy to verify that the three occupied nodes of
Fig.~\ref{fig:MDSexample}B form a MDS for that small network.
Notice a network may have more than one MDS.

\subsection{The GLR Process}

Here we extend the leaf-removal idea of \cite{Bauer-Golinelli-2001} (see
also more recent work 
\cite{Echenique-etal-2005,Lucibello-RicciTersenghi-2014,Takabe-Hukushima-2014}) 
and consider a generalized leaf-removal process. This dynamics is based on 
the following two considerations: 
first, as pointed out in \cite{Bauer-Golinelli-2001,Echenique-etal-2005},
if node $i$ is an unobserved leaf node (which has only a single neighbor,
say $j$), then occupying $j$ but leaving $i$ empty must be an optimal
strategy; second, we notice that if $i$ is an empty but observed node
and at most one of its adjacent nodes is unobserved, then it must be an 
optimal strategy \emph{not} to occupy $i$. This second
point was not considered in the conventional leaf-removal
process \cite{Echenique-etal-2005}.

\begin{figure}
  \begin{center}
    \includegraphics[width=0.5\textwidth]{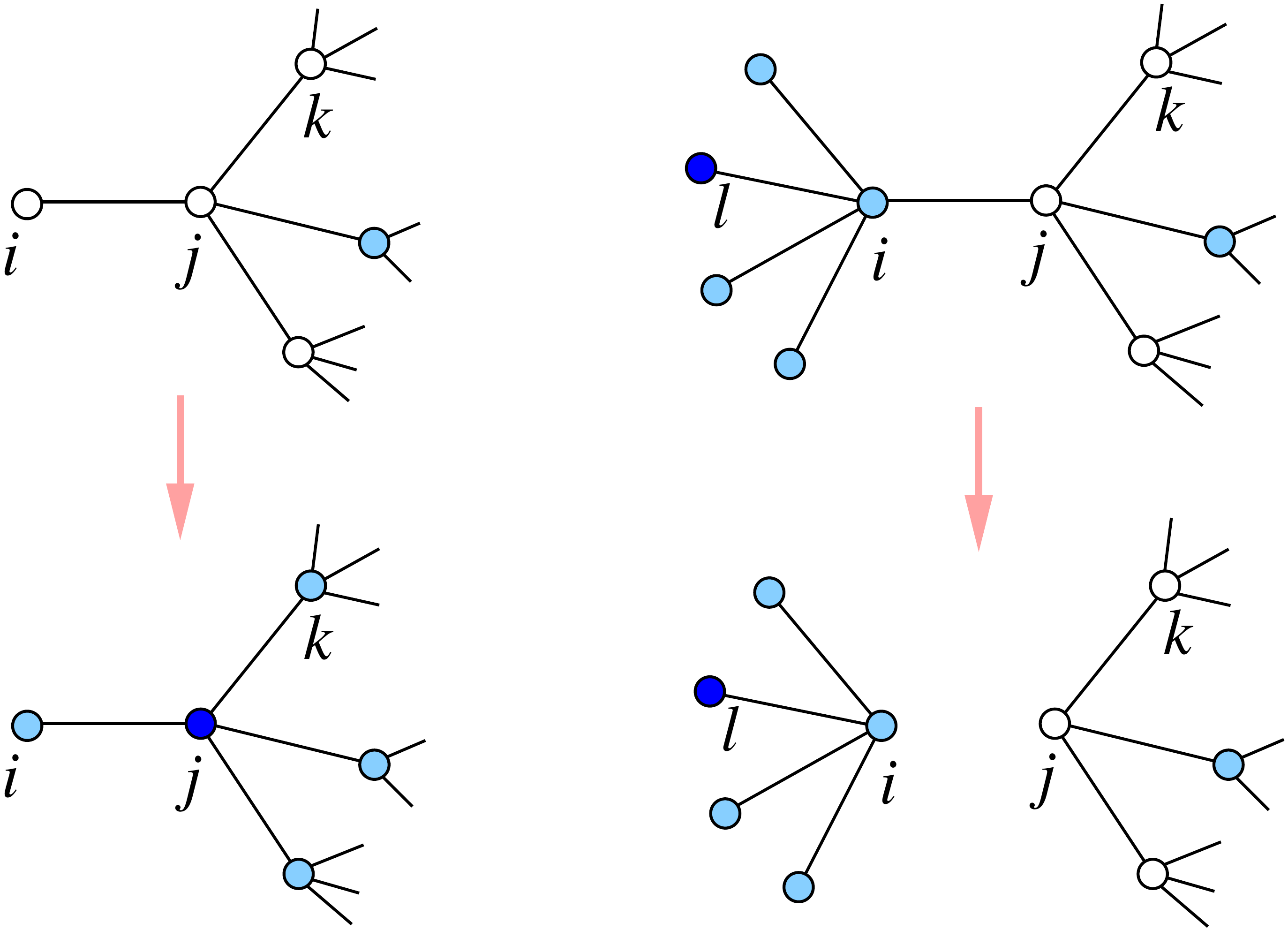}
  \end{center}
  \caption{
    \label{fig:GLRidea}
    The two basic operations of the generalized leaf-removal
    process. White circles denote
    empty and unobserved nodes, cyan (light gray) circles denote 
    empty but observed nodes, and blue (dark gray) circles denote 
    occupied nodes. (Left panel) the unique
    adjacent node $j$ of an unobserved leaf node $i$ is occupied, and
    all the neighbors of $j$ are observed.
    (Right panel) an empty observed node $i$ has only a single unobserved 
    neighbor $j$, then the link between $i$ and $j$ is deleted.
  }
\end{figure}

The GLR process simplifies the input network $W$ at discrete 
evolution steps $t=0, 1, 2, \ldots$.
For the convenience of description, let us denote 
by $W_t$ the simplified network at the start of the $t$-th
evolution step of GLR. $W_0$ at the initial
step $t=0$ is identical to the original network $W$, and all the nodes of
$W_0$ are unobserved. 
We prove that if GLR makes the whole input network $W$ be observed, 
then the set of nodes occupied during this process must be a MDS. For this
latter purpose, let us denote by $\Gamma_0$ a 
MDS of the input network $W$ (there must be at least one such set). 
The essential idea is to demonstrate that during GLR,
we can modify $\Gamma_{0}$ in such a way that its size does not change but
all the nodes $i$ that are fixed to be occupied ($c_i=1$) are in 
$\Gamma_{0}$ while all the nodes $j$ that are fixed to be unoccupied 
($c_j=0$) are not in $\Gamma_{0}$.
Starting from evolution step $t=0$, let us perform GLR and 
modify $\Gamma_{0}$ in the following sequential order:
\begin{enumerate}
\item[(0)]
  As long as there is an isolated node $i$ in network $W_t$,
  fix its occupation state to $c_i=1$ and delete it from $W_t$. 
  All such fixed nodes $i$ must also belong to $\Gamma_{0}$.
  
\item[(1)]
  As long as there is a leaf node $i$ in network $W_t$ which is
  not yet observed, fix the occupation state of its unique neighbor
  $j$ to $c_j=1$ and fix that of $i$ to $c_i=0$ so that $j$ and all its
  adjacent nodes (including $i$) are now observed, see
  Fig.~\ref{fig:GLRidea} (left panel).
  We then delete node $j$ and all its connected links from $W_t$.
  If $j$ belongs to  $\Gamma_{0}$ then node $i$ must not belong 
  to it, because otherwise $\Gamma_{0}$ could not have been a MDS. 
  On the other hand, if node $j$ does not belong
  to $\Gamma_{0}$ then node $i$ must belong to it, and in this latter
  case we modify $\Gamma_{0}$ by adding $j$ to it and deleting
  $i$ from it.
  
\item[(2)]
  Then as long as there is a node $i$ which is itself observed and which has
  only a single unobserved neighbor $j$, delete the link 
  $(i,j)$ from network $W_t$, see Fig.~\ref{fig:GLRidea} (right panel).
  We do not modify $\Gamma_{0}$ if
  node $i$ does not belong to it.
  If node $i$ does belong to $\Gamma_{0}$ then node $j$ must 
  not belong to it, and in this latter case
  we add $j$ to $\Gamma_{0}$ and delete $i$ from it.
  
\item[(3)]
  Then as long as there is an observed node $i$ which is not connected to
  any unobserved node, fix its occupation state to $c_i=0$ and
  delete it and all its attached links from $W_t$. 
  Such a node $i$ must not belong to $\Gamma_{0}$, for
  otherwise $\Gamma_{0}$ could not have been a MDS.
  
\item[(4)]
  If the resulting network $W_t$ is empty or it contains no isolated node
  nor leaf node, the GLR process stops. 
  If  $W_t$ still contains at least one isolated or 
  leaf node, then we increase the evolution step from $t$ to $(t+1)$ 
  and initialize the network $W_{t+1}$ as identical to $W_t$. A node $i$ of
  $W_{t+1}$ is regarded as observed if and only if 
it is observed in network $W_t$.
  We then repeat the above-mentioned operations (1)--(3).
\end{enumerate}

If the final simplified network is non-empty, then there must be some nodes 
that are still unobserved after the GLR process. The subnetwork induced
by these unobserved nodes is referred to as the \emph{core} of the original 
network $W$. This core is connected only to observed empty nodes but not to
occupied nodes. We denote by $n_{core}$ the fraction of nodes in this core and
by $w$ the fraction of occupied nodes. 

If the original network $W$ has no core, 
then the set $\Gamma$ of occupied nodes by the GLR 
process must be identical to the final $\Gamma_{0}$, which is a MDS
modified from the original MDS. We have therefore proven that GLR constructs
a MDS for a network $W$ if this network contains no core.
(All the above-mentioned modification operations on $\Gamma_0$ are
ignored in the actual implementation of the GLR process. They are introduced
here solely for proving
 that GLR is able to construct a MDS if there is no core.)
Furthermore, we notice that if the GLR process 
finishes with some nodes remaining to be unobserved, the set of
nodes occupied during this process must be a MDS for the 
subnetwork of $W$ induced by  all the observed nodes. 
This is because all these occupied nodes also belong to the 
modified MDS $\Gamma_{0}$, while all those nodes fixed to be 
unoccupied are outside of $\Gamma_{0}$.

\begin{figure}
  \begin{center}
    \includegraphics[angle=270,width=0.6\textwidth]{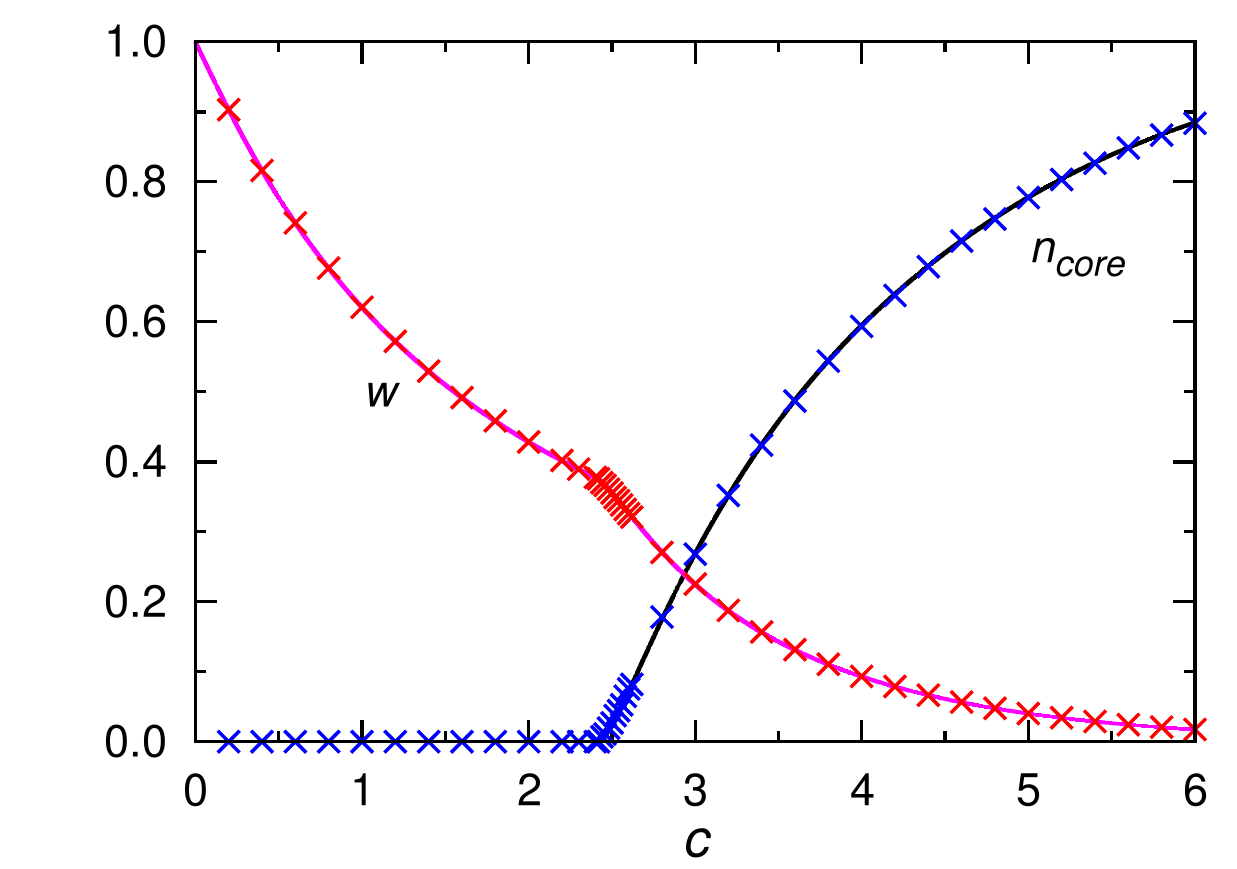}
  \end{center}
  \caption{ \label{fig:erglr}
    Generalized leaf-removal on Erd\"os-R\'enyi random networks.
    $w$ and $n_{core}$ are the fractions of occupied and 
    unobserved nodes, respectively.
    Cross symbols are results obtained by running GLR on a single
    ER network of size $N=10^6$ and mean degree $c$; solid lines
    are the predictions of the percolation theory for $N=\infty$.
  }
\end{figure}

\begin{figure}
  \begin{center}
    \includegraphics[angle=270,width=0.6\textwidth]{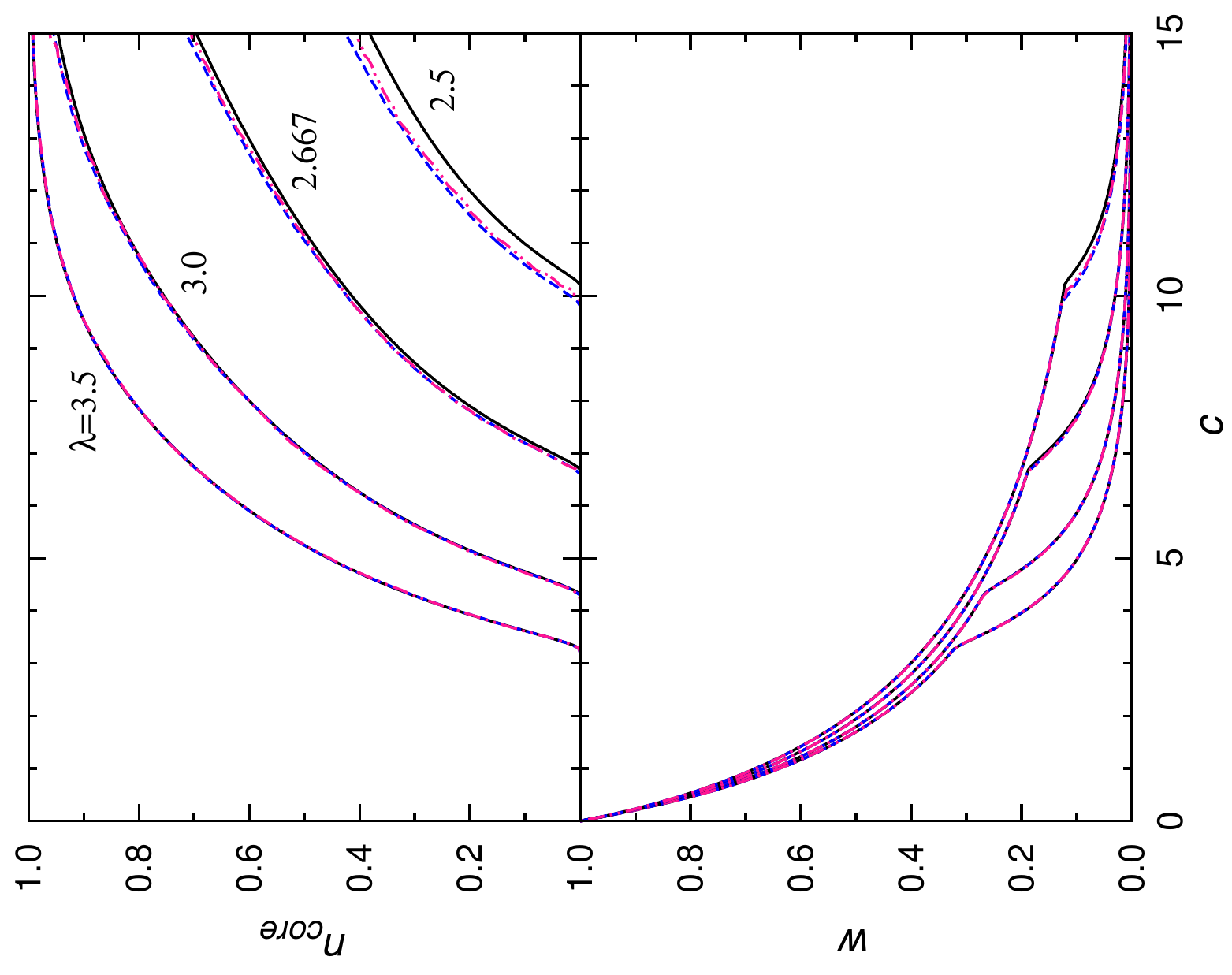}
  \end{center}
  \caption{ \label{fig:sfstaticglr}
    Generalized leaf-removal on scale-free random networks
    of decay exponent $\lambda=3.5, 3.0, 2.667, 2.5$ (from left to right)
    generated through the static model \cite{Goh-Kahng-Kim-2001}. 
    $w$ and $n_{core}$ are the fractions of occupied and 
    unobserved nodes, respectively.
    Red dash-dotted lines are results 
    obtained by running GLR on a single network
    instance of size $N=10^6$ and mean degree $c$, while blue dashed lines
    are results obtained by the core percolation theory using the degree profile
    of this network instance as input. Black solid lines are the 
    predictions of the percolation theory for $N=\infty$.
  }
\end{figure}

\begin{figure}
  \begin{center}
    \includegraphics[angle=270,width=0.6\textwidth]{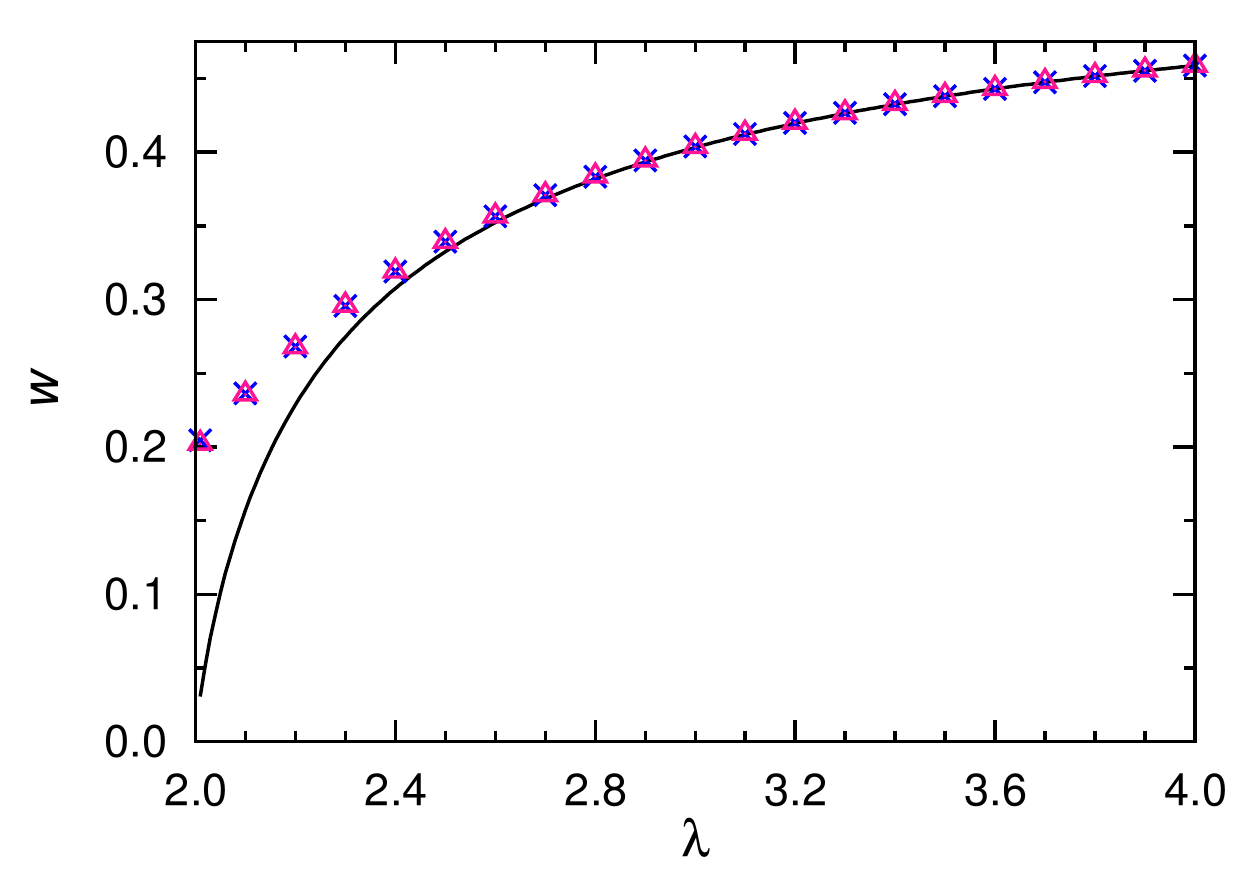}
  \end{center}
  \caption{ \label{fig:sfpureglr}
    Generalized leaf-removal on pure scale-free random networks with 
    minimum degree $d=1$. $w$ is the fraction of occupied nodes. The
    fraction $n_{core}$ of unobserved nodes is simply $n_{core}=0$.
    Red triangle symbols are results obtained by running GLR on a single
    pure SF network of size $N=10^6$ and  decay exponent
    $\lambda$, while blue cross symbols are results obtained by the 
    percolation theory using the degree profile
    of this network instance as input. The black solid line is obtained
    by the percolation theory at $N=\infty$.
  }
\end{figure}

We generate many large instances of Erd\"os-R\'enyi (ER) and scale-free
(SF) random networks and run the GLR process on them 
(details of the network
sampling method are given in
sections~\ref{sub:coreer}, \ref{sub:coresfstatic}, and
\ref{sub:coresfpure}). Some representative results are shown in
Fig.~\ref{fig:erglr} for ER networks
\cite{He-Liu-Wang-2009,Albert-Barabasi-2002}, 
in Fig.~\ref{fig:sfstaticglr} for
SF networks generated through the static model \cite{Goh-Kahng-Kim-2001},
and in Fig.~\ref{fig:sfpureglr} for pure SF networks 
\cite{He-Liu-Wang-2009,Albert-Barabasi-2002}. A major observation is that
there is no core in pure SF random networks with minimum node degree
$d_{min}=1$, therefore a MDS for such a network can be easily constructed by 
the GLR process. Another major observation is that there is a 
continuous core percolation
transition in ER networks and in SF networks generated through the static 
model. This core percolation transition occurs at certain threshold value of 
the mean node degree.
For example, for ER networks with 
$N=10^6$ nodes and $M= (c/2) N$ links, when the mean node degree $c<2.41$ 
there is no core ($n_{core}=0$), and GLR reaches a MDS for the whole 
network (Fig.~\ref{fig:erglr}). The core emerges at $c\approx 2.41$ and
its relative size $n_{core}$ then increases with $c$ continuously from zero.
For $c>2.41$, GLR constructs a MDS only for part of the 
ER network and it leaves an extensive core of $n_{core} N$ unobserved nodes. 

Notice the core percolation transition resulting from the GLR optimization
process is qualitatively different from the simpler observability
transition discussed in \cite{Yang-Wang-Motter-2012}, 
which considers the appearance of a giant connected component of
observed nodes resulting from an initial set of randomly chosen
occupied nodes. We now develop a percolation theory to 
thoroughly understand the GLR dynamics on random
networks.

\subsection{The Core Percolation Theory}

A random network is characterized by a degree distribution $P(d)$, 
which gives the
fraction of nodes with degree $d\geq 0$ \cite{He-Liu-Wang-2009}.
We assume that there is no correlation between the degrees of adjacent
nodes, therefore the degree $d$ of a node reached by following a 
randomly chosen link obeys the distribution $Q(d)$ of the form
\begin{equation}
Q(d) \equiv  \frac{ P(d) d}{c} \; , 
\end{equation}
where $c\equiv \sum_{d \geq 0} P(d) d$ 
is the mean node degree of the network. 
Consider a link $(i,j)$ of the network $W$. Let us neglect for the
moment the constraint of node $i$ to node $j$ but only consider the other
adjacent nodes of $j$. If the constraint of node $i$ is
neglected, then what is
the probability $\alpha_t$ that node $j$ becomes an unobserved leaf node
(i.e., it has no other adjacent node except $i$) at the start of the
$t$-th GLR evolution step? What is the probability
$\beta_t$ that $j$ becomes newly occupied ($c_j=1$) at the $t$-th GLR
step? What is the probability $\gamma_t$ that $j$ is observed but not
occupied at the end of the $t$-th GLR step? And what is the probability
$\eta_t$ that at the end of the $t$-th GLR step, node $j$ is an
observed and unoccupied node and it has no unobserved adjacent node except
 $i$? For an uncorrelated random network these four sets of probability
parameters $\{\alpha_0, \alpha_1, \ldots\}$, 
$\{\beta_0, \beta_1, \ldots \}$, $\{\gamma_0, \gamma_1, \ldots \}$, 
and $\{\eta_0,\eta_1, \ldots\}$ can be computed by a set of iterative 
equations.

The expressions of $\alpha_0, \beta_0, \gamma_0, \eta_0$
for the initial evolution step $t=0$ are
\begin{eqnarray}
  \alpha_0 &=& Q(1) \; , \label{eq:alpha0}\\
  \beta_0 &=& 1- \sum\limits_{d\geq 1} Q(d) (1-\alpha_0)^{d-1}  \; , 
  \label{eq:beta0} \\
  \gamma_0 &=& \sum\limits_{d\geq 1} Q(d)
  \bigl[ (1-\alpha_0)^{d-1} - (1-\alpha_0 - \beta_0)^{d-1} \bigr] \; , 
\label{eq:gamma0}\\
  \eta_0 &=& \sum\limits_{d\geq 1} Q(d) \bigl[
    (\beta_0 + \gamma_0)^{d-1} - \gamma_0^{d-1} \bigr] \; .
  \label{eq:eta0}
\end{eqnarray}
Equation (\ref{eq:alpha0}) is trivial, it simply describes the situation
that node $j$ has only a single neighbor (i.e., node $i$).
Equation (\ref{eq:beta0}) describes the situation that node $j$ is adjacent
to at least one leaf node (except $i$), which will
guarantee $j$ to be occupied at the $t=0$ GLR step.  A random network has 
only very few short loops and therefore the local network structure around
node $j$ is a tree. In the core percolation theory we therefore assume that
the adjacent nodes of $j$ are completely independent of
each other when $j$ is still unobserved (such an assumption was also
exploited in our earlier percolation studies
\cite{Zhao-Zhou-Liu-2013,Zhou-2005a,Zhou-2012}).
Based on this assumption, the probability of all the 
adjacent nodes (except $i$) of $j$ not being unobserved 
leaves is then written in Eq.~(\ref{eq:beta0}) as the product of the
individual probability $(1-\alpha_0)$ of an adjacent node not being an 
unobserved leaf.
Equation (\ref{eq:gamma0}) expresses the fact that for node $j$ to be
an unoccupied but observed node at the end of the $t=0$ evolution step,
it should not be adjacent to any unobserved leaf node but 
at least one of its adjacent nodes (except $i$)
should be occupied.

If node $j$ is adjacent to one or more nodes that are occupied at the
$t=0$ evolution step and all its other adjacent nodes (except $i$)
are observed at this evolution step, then at the end of this evolution 
step $j$ is unoccupied but observed and it is not adjacent to any
unobserved node (except $i$). This then leads to the expression 
(\ref{eq:eta0}) for $\eta_0$. 
Notice such an observed but unoccupied node $j$ will be deleted at the end
of the $t=0$ evolution step. After all such nodes are deleted, 
some unobserved nodes in the remaining network may become isolated or
be connected to only a single node.
If this is the case, these isolated or leaf nodes will trigger the
next ($t=1$) evolution step.

Following the same line of theoretical considerations, we obtain the
expressions of $\alpha_t$, $\beta_t$, $\gamma_t$, and $\eta_t$ for
the $t$-th GLR evolution step ($t\geq 1$) as
\begin{eqnarray}
  \alpha_t &=& \left\{
\begin{array}{ll}
  \sum\limits_{d\geq 1} Q(d) ( \eta_0 )^{d-1} - Q(1) \; ,  & \quad \quad (t=1) \\
  \sum\limits_{d\geq 1} Q(d) \Bigl[ 
    \bigl( \sum\limits_{l=0}^{t-1}
    \eta_l \bigr)^{d-1} -
    \bigl( \sum\limits_{l=0}^{t-2} \eta_l \bigr)^{d-1}
    \Bigr] \; , & \quad \quad (t\geq 2) 
\end{array}\right.
  \label{eq:alphat}\\
  \beta_t  &=& \sum\limits_{d\geq 1} Q(d) \Bigl[ 
    \bigl(1-\sum\limits_{l=0}^{t-1}
    \alpha_l \bigr)^{d-1} - \bigl(1- \sum\limits_{l=0}^{t} \alpha_l \bigr)^{d-1}
    \Bigr] \; ,
  \label{eq:betat}\\
  \gamma_t &=& \sum\limits_{d\geq 1} Q(d) \Bigl[ \bigl(1-\sum\limits_{l=0}^{t}
    \alpha_l \bigr)^{d-1} - \bigl(1-\sum\limits_{l=0}^{t} (\alpha_l + \beta_l)
    \bigr)^{d-1} \Bigr] \; ,
  \label{eq:gammat}\\
  \eta_t   &=& \sum\limits_{d\geq 1} Q(d) \Bigl[ \bigl( \sum\limits_{l=0}^{t}
    \beta_l + \gamma_t\bigr)^{d-1} - (\gamma_t )^{d-1} \Bigr]
  -\sum\limits_{l=0}^{t-1} \eta_l \; .
  \label{eq:etat}
\end{eqnarray}

Let us denote by $\gamma_{lim}$ the value of $\gamma_{t}$ at the
last evolution step $t=t_{lim}$ of the GLR process (notice that the
maximal evolution step $t_{lim}$ may approach infinity for a
network with $N=\infty$ nodes).
Furthermore, we define the accumulated values of $\alpha_t$, $\beta_t$, 
and $\eta_t$ as
$$
  \alpha_{cum} \equiv  \sum\limits_{t\geq 0} \alpha_t \; , \quad
  \beta_{cum} \equiv  \sum\limits_{t\geq 0} \beta_t \; , \quad
  \eta_{cum} \equiv  \sum\limits_{t\geq 0} \eta_t \; .
$$
There are the following relationships among $\alpha_{cum}$,
$\beta_{cum}$, $\eta_{cum}$ and $\gamma_{lim}$:
\begin{eqnarray}
  \alpha_{cum} &= & \sum\limits_{d\geq 1} Q(d) (\eta_{cum})^{d-1} \; ,
  \label{eq:alphacum} \\
  \beta_{cum} &= & 1- \sum\limits_{d\geq 1} Q(d) (1-\alpha_{cum})^{d-1} \; ,
  \label{eq:betacum}\\
  \gamma_{lim} &=& \sum\limits_{d\geq 1} Q(d) \Bigl[
    (1-\alpha_{cum})^{d-1} - (1- \alpha_{cum}- \beta_{cum})^{d-1} \Bigr] \; ,
  \label{eq:gammacum}\\
  \eta_{cum} &=& \sum\limits_{d\geq 1} Q(d) \Bigl[
    (\beta_{cum} + \gamma_{lim})^{d-1} - (\gamma_{lim})^{d-1} \Bigr] \; .
  \label{eq:etacum}
\end{eqnarray}

After all the probability parameters $\alpha_t$,
$\beta_t$, $\gamma_t$, $\eta_t$ (for $t=0, 1, \ldots$) for a node
$j$ at the end of a link $(i, j)$ are determined 
by neglecting the constraint associated with node $i$, we now ask the 
following two questions: If the constraint
of node $i$ to all its adjacent nodes are considered, then what is the 
probability $n_{core}$ of $i$ to be unobserved after the whole GLR 
process? And what is the probability $I_t$ of node $i$ to
be occupied at the $t$-th GLR evolution step?
If node $i$ remains to be unobserved during the whole GLR process, it must
not be adjacent to any unobserved leaf node nor to any occupied node, and
it must have at least two adjacent nodes after the whole GLR process. 
Therefore we obtain that
\begin{eqnarray}
n_{core}  &=& \sum\limits_{d\geq 2} P(d) \sum\limits_{s=0}^{d-2}
C_{d}^{s} (\eta_{cum})^{s} (1-\alpha_{cum}-\beta_{cum} - \eta_{cum})^{d-s} 
\nonumber \\
&=& \sum\limits_{d\geq 1} P(d) \bigl[
  (1-\alpha_{cum}-\beta_{cum})^{d}-(\eta_{cum})^{d} \nonumber \\
& & \quad \quad \quad \quad \quad 
  - d (\eta_{cum})^{d-1} (1-\alpha_{cum}-\beta_{cum}-\eta_{cum}) \bigr] \; ,
\label{eq:ncore}
\end{eqnarray}
where $C_{d}^{s} \equiv d!/[s! (d-s)!]$ is the binomial coefficient.
Notice that if $(\alpha_{cum}+\beta_{cum} + \eta_{cum})=1$ then we
have $n_{core}=0$.

It is easy to see that the probability $I_0$ of a randomly chosen
node $i$ to be occupied  at the $t=0$ GLR evolution step is
\begin{equation}
  \label{eq:I0}
  I_0 = 1 - P(1) (1- \frac{\alpha_0}{2}) - 
  \sum\limits_{d\geq 2} P(d) (1-\alpha_0)^{d} \; .
\end{equation}
The coefficient $1/2$ in the second term of the above expression
reflects the fact that if node $i$ has only one neighbor $j$, then
$i$ has one-half probability to be occupied if $j$ also has only
one neighbor (namely $i$).

If a randomly chosen node $i$ is not occupied at
the $t=0$ evolution step, then the 
probability $I_1$ of it being occupied at the $t=1$ evolution step is
\begin{eqnarray}
  I_1 & = &  \sum\limits_{d\geq 2} P(d) ( \eta_0 )^{d} +
  \sum\limits_{d\geq 2} P(d) \Bigl[
    (1-\alpha_0)^{d} - (1-\alpha_0 - \alpha_1)^{d} \nonumber \\
& & \quad 
    - d \alpha_1 \bigl(  (\beta_0+\gamma_0)^{d-1} - (\gamma_0)^{d-1} \bigr)
\Bigr] 
  -\frac{1}{2} \sum\limits_{d\geq 2} P(d) d \alpha_1 (\eta_0)^{d-1} \; .
 \label{eq:I1a}
\end{eqnarray}
All the adjacent nodes of  $i$ might haven been deleted at the end of the
$t=0$ evolution step. If this is the case node $i$ becomes
isolated at the start of the $t=1$ evolution step, which leads to the
first summation of Eq.~(\ref{eq:I1a}).
The second summation of Eq.~(\ref{eq:I1a}) corresponds to the other
situation of node $i$ not being occupied nor being deleted at the 
$t=0$ evolution step but it is adjacent to at least one node that becomes
an unobserved leaf at the start of the $t=1$ evolution step.
Notice if node $i$ becomes an unobserved leaf node at the start of the
$t=1$ evolution step with its unique neighbor also being such a leaf node,
then $i$ has only one-half probability to be occupied at this evolution
step. This last situation leads to the third summation
term of Eq.~(\ref{eq:I1a}), which corrects the over-counted probability of
occupation in the second summation term.

Following the same line of theoretical considerations, we obtain the
probability $I_t$ of a randomly chosen node $i$ changing from being unoccupied
to being occupied at the $t$-th GLR evolution step ($t\geq 2$):
\begin{eqnarray}
  I_t &= &  \sum\limits_{d\geq 2} P(d) \Bigl[ \bigl(\sum\limits_{l=0}^{t-1}
    \eta_l \bigr)^{d} - \bigl( \sum\limits_{l=0}^{t-2} \eta_l \bigr)^{d}
    - d \eta_{t-1}  \bigl(\sum\limits_{l=0}^{t-2} \eta_l \bigr)^{d-1} \Bigr] 
  \nonumber \\
  & & 
  + \sum\limits_{d\geq 2} P(d) \Bigl[ \bigl(1-\sum\limits_{l=0}^{t-1}
    \alpha_l \bigr)^{d} - \bigl(1-\sum\limits_{l=0}^{t} \alpha_l \bigr)^{d} \Bigr]
  \nonumber \\
  & & - \sum\limits_{d\geq 2} P(d) d 
  \alpha_t \Bigl[ \bigl(\sum\limits_{l=0}^{t-1} \beta_l + \gamma_{t-1} \bigr)^{d-1}
    - (\gamma_{t-1})^{d-1} +
    \bigl(\sum\limits_{l=0}^{t-2} \eta_l \bigr)^{d-1} \Bigr]
  \nonumber \\
  & & - \frac{1}{2} \sum\limits_{d\geq 2} P(d) d \alpha_{t}
  \Bigl[ \bigl( \sum\limits_{l=0}^{t-1} \eta_l \bigr)^{d-1}
    - \bigl(\sum\limits_{l=0}^{t-2} \eta_l \bigr)^{d-1} \Bigr] \; .
\end{eqnarray}
The probability $w$ of a randomly chosen node $i$ to be occupied during
the GLR process is then
\begin{eqnarray}
  w &= & \sum\limits_{t\geq 0} I_t \label{eq:wa} \\
  & = & 1 - P(1) (1- \alpha_0 / 2) - \sum\limits_{d\geq 2} P(k) 
  \bigl[ (1- \alpha_{cum})^d - (\eta_{cum})^{d} \bigr]
 \nonumber \\
 & &  - \sum\limits_{t\geq 1} \sum\limits_{d\geq 2} P(d) d \Bigl[
   \eta_t  \bigl( \sum\limits_{l=0}^{t-1} \eta_l \bigr)^{d-1} 
   + \alpha_t \bigl(
   \sum\limits_{l=0}^{t-1} \beta_l + \gamma_{t-1} \bigr)^{d-1}
   -  \alpha_t (\gamma_{t-1})^{d-1} \Bigr] \nonumber \\
   & & - \frac{1}{2} \sum\limits_{t\geq 2} \sum\limits_{d\geq 2}
 P(d) d \alpha_t \Bigl[ \bigl( \sum\limits_{l=0}^{t-1} \eta_l \bigr)^{d-1}
   + \bigl( \sum\limits_{l=0}^{t-2} \eta_l \bigr)^{d-1} \Bigr]  \nonumber \\
 & & - \frac{1}{2} \sum\limits_{d\geq 2} P(d) d \alpha_1 (\eta_0)^{d-1}
 \; .
 \label{eq:wb}
\end{eqnarray}

Our core percolation theory can be applied both to single finite 
random network
instances and to random network ensembles at the thermodynamic limit
$N\rightarrow \infty$. 
For each $t$ (starting from $t=0$), we first compute $\alpha_t$, 
then use $\alpha_t$ as input to compute
$\beta_t$, then use $\alpha_t$ and $\beta_t$ as inputs to
compute $\gamma_t$, and finally use $\alpha_t$, $\beta_t$ and
$\gamma_t$ as inputs to compute $\eta_t$. 
For a finite random network of $N$ nodes, the iteration stops if the 
evolution step $t$ increases to a value $t_{lim}$ such that 
$I_{t_{lim}} < 1/N$. This is because if $N I_{t} < 1$ the number of 
newly occupied nodes has a high probability to be zero and
then GLR will be unable to continue.
For the case of $N\rightarrow \infty$, the numerical iteration process
can be carried out to a sufficiently large evolution step $t=t_{lim}$
until $\alpha_{t_{lim}} \approx 0$.

\subsection{Results on Erd\"os-R\'enyi Random Networks}
\label{sub:coreer}

We generate an ER random network $W$ of $N$ nodes and $M=(c/2) N$ links by
adding links sequentially to an initial network of $N$ isolated nodes. To add
a link, we choose two different nodes $i$ and $j$ uniformly at random 
from the whole node set and set up a link $(i, j)$ between them
if this link has not been created before. The mean node degree of the
resulting network $W$ is equal to $c$. When the number $N$ of nodes
is sufficiently large the degree distribution $P(d)$ of such a
ER network obeys the Poisson distribution
\cite{He-Liu-Wang-2009,Albert-Barabasi-2002}
\begin{equation}
  P(d)= \frac{c^d e^{-c}}{d!} \quad \quad \quad \quad (d \geq 0) \; .
\end{equation}

For this network ensemble, the predicted results of 
$n_{core}$ and $w$ by our core
percolation theory are in perfect agreement with
simulation results (see Fig.~\ref{fig:erglr}). Especially, at
the thermodynamic limit $N\rightarrow \infty$, the theory
predicts a continuous core percolation phase
transition at $c\approx 2.4102$, which 
is slightly lower than the core percolation phase transition
point of $c\approx 2.7183$ caused by the conventional
leaf-removal process \cite{Bauer-Golinelli-2001}.
Before the GLR-induced  core percolation transition occurs,
the occupation fraction $w$ obtained by Eq.~(\ref{eq:wb}) is 
equal to the ensemble-averaged MDS size (relative to $N$), but it is only a 
lower bound to this size when an extensive core emerges in the
random network ($n_{core}>0$).

\subsection{Results on Scale-free Random Networks Generated through the
Static Model}
\label{sub:coresfstatic}

Now let us consider GLR-induced
core percolation on more heterogeneous random networks.
We generate a scale-free network $W$ of $N$ nodes and $M=(c/2) N$ links
according to the static model \cite{Goh-Kahng-Kim-2001}. Each node
$i \in \{1, 2, \ldots, N\}$ is first assigned a 
fitness value $\theta_i = i^{-\xi}/(\sum_{j=1}^{N} j^{-\xi})$,
where $ 0 \leq \xi < 1$ is a control parameter. Then we add links
between pairs of these $N$ nodes in a sequential manner.
To create a link, two nodes $i$ and $j$ are chosen independently 
from the set of $N$ nodes, and the 
probability that $i$ and $j$ being chosen is equal to
$\theta_i \theta_j$; if nodes $i$ and $j$ are different and the 
link $(i, j)$ has not been created before,  this link is added to
network. The final network $W$
has a power-law degree distribution $P(d) \propto d^{-\lambda}$ for $d \gg 1$,
with degree decay exponent $\lambda = 1 + 1/ \xi $. In the thermodynamic
limit $N\rightarrow \infty$, an explicit expression for $P(d)$ is
obtained as \cite{Catanzaro-PastorSatorras-2005}
\begin{equation}
  P(d)=\frac{[c (1-\xi)]^d}{d! \xi }
  \int_{1}^{\infty} {\rm d} x e^{-c (1-\xi) x} x^{k-1-1/\xi} 
\quad \quad \quad \quad (d\geq 0) \; .
\end{equation}

\begin{figure}
  \begin{center}
    \includegraphics[angle=270,width=0.6\textwidth]{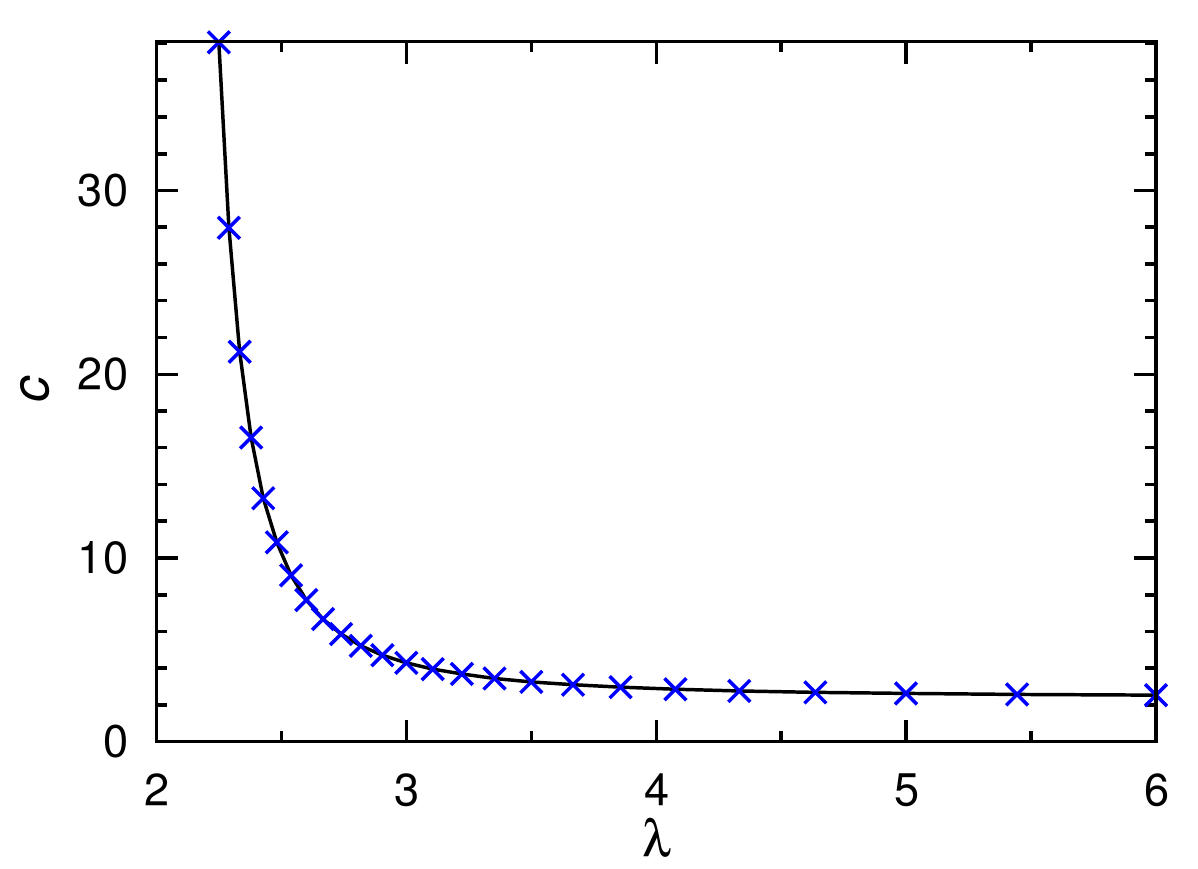}
  \end{center}
  \caption{ \label{fig:coretranSFstatic}
    Core percolation phase transition in infinite ($N=\infty$)
    random scale-free networks generated through
    the static model \cite{Goh-Kahng-Kim-2001}.
    Horizontal axis is the degree decay exponent $\lambda$, while
    vertical axis is the value of the mean node degree $c$ at the
    phase transition point. Cross symbols are predictions of the
    core-percolation theory, while the solid line is just a guide for the
    eye.
  }
\end{figure}

For $N=\infty$, a continuous core percolation phase transition is observed
in such a SF network, and
this transition occurs at more and more larger value of
the mean node degree $c$ as the decay exponent $\lambda$ decreases
(see Fig.~\ref{fig:sfstaticglr} for
$\lambda=3.5$, $3.0$, $8/3\approx 2.667$, and $2.5$ and
Fig.~\ref{fig:coretranSFstatic} for $2< \lambda \leq 6$).
When the decay
exponent $\lambda$ is less then $3.0$, theoretical predictions obtained 
at $N=\infty$  are quantitatively different from theoretical
and simulation results obtained on finite (e.g., $N=10^6$)
network instances, with the deviations become more
pronounced as $\lambda$ is closer to $2.0$. Such a finite-size effect is
mainly caused by the natural cutoff of maximum node degree in 
finite networks (it was also observed in our earlier work
\cite{Zhao-Zhou-Liu-2013} on another type of percolation transitions).
We emphasize that for a give finite value of $N$, the results of
the core percolation theory agree with the simulation results of the
actual GLR process very well, especially if we average the theoretical
and simulation results over many network instances
to reduce fluctuations.

For random SF networks generated through the static model with $N=\infty$ nodes,
the core percolation transition value of
mean node degree $c$ is very sensitive 
to the decay exponent $\lambda$ in the region of
$2 < \lambda < 2.5$, and it diverges as 
$\lambda$ approaches $2.0$ from above (Fig.~\ref{fig:coretranSFstatic}).
At the other limit of $\lambda\rightarrow \infty$, the mean node degree
at the phase transition approaches the  value of $c\approx 2.4102$, which
is just the core percolation phase transition point of an infinite
ER random network.

\subsection{Results on Pure Scale-Free Random Networks}
\label{sub:coresfpure}

When $N=\infty$, a pure scale-free random network has the following
degree distribution
\begin{equation}
P(d) = \frac{1}{\sum_{k=1}^{\infty} k^{-\lambda}} d^{-\lambda}
\quad \quad \quad \quad (d \geq 1) \; ,
\end{equation}
with $\lambda > 2$ to ensure a finite value for the mean node degree $c$.
For such a random SF network our core percolation theory predicts that
$n_{core}=0$, namely there is no core percolation transition and the GLR
process will construct a MDS for the whole network. The fraction $w$ of
occupied nodes (i.e., the size of a MDS relative to the node number $N$)
decreases with the decreasing of the degree decay exponent $\lambda$ (see 
Fig.~\ref{fig:sfpureglr}), and it approaches zero as $\lambda$ approaches
$2.0$ from above.

We also generate a set of pure SF random networks of finite size $N$ following
the same procedure as mentioned in \cite{Zhou-Lipowsky-2005} (see also the
supplementary information of \cite{Zhao-Zhou-Liu-2013}). The minimum node
degree of such a SF network is $d_{min}=1$, while the maximum node degree
is $d_{max}\approx N^{1/(\lambda -1)}$ \cite{Zhou-Lipowsky-2005}.
When we apply both the GLR process and the core percolation theory on
these finite network instances, we find the simulation results 
on the fraction $n_{core}$ of nodes in the core
and the fraction $w$ of occupied nodes are in perfect agreement with
the corresponding theoretical results (see Fig.~\ref{fig:sfpureglr}).
All these finite SF networks contain no core ($n_{core}=0$), and the MDS
relative size $w$ is an increasing function of $\lambda$.

Figure~\ref{fig:sfpureglr} also demonstrates strong finite-size effect
for pure SF random networks of $\lambda<3.0$. This finite-size effect
is again mainly caused by the cutoff of the maximum node degree of finite
networks, which makes the mean node degree of a finite network be
smaller than that of an infinite network. For example, at $\lambda=2.1$ the
mean node degree of an infinite network is $c\approx 61.49$, while
that of a finite network of size $N=10^6$ is reduced to $c\approx 5.134$.

\section{Hybrid Local Algorithm}
\label{sec:hybrid}

There is a very simple greedy algorithm in the literature to solve the
MDS problem approximately, which is based on the concept of node impact
\cite{Haynes-Hedetniemi-Slater-1998,Takaguchi-Hasegawa-Yoshida-2014,Molnar-etal-2013}).
The impact of an unoccupied node $i$ equals to the number of nodes that will
be observed by occupying $i$. For example, if node $i$ has
$3$ unobserved neighbors, its impact is $4$ if $i$ is itself unobserved 
and is $3$ if $i$ is already adjacent to one or more
occupied nodes.
Starting from an input network $W$ with all the nodes
unobserved, the greedy algorithm selects uniformly at random a node
$i$ from the subset of nodes with the highest impact and fix its occupation
state to $c_i=1$. All the adjacent nodes of $i$ are then
observed. If there are still unobserved nodes in the network, the impact
value for each of the unoccupied nodes is updated and the greedy occupying
process is repeated until all the nodes are observed.
This pure greedy algorithm is very easy to implement and very fast,
but we find that it usually fails to reach a true MDS even when the 
input network contains no core.

\begin{figure}
  \begin{center}
    \includegraphics[angle=270,width=1.0\textwidth]{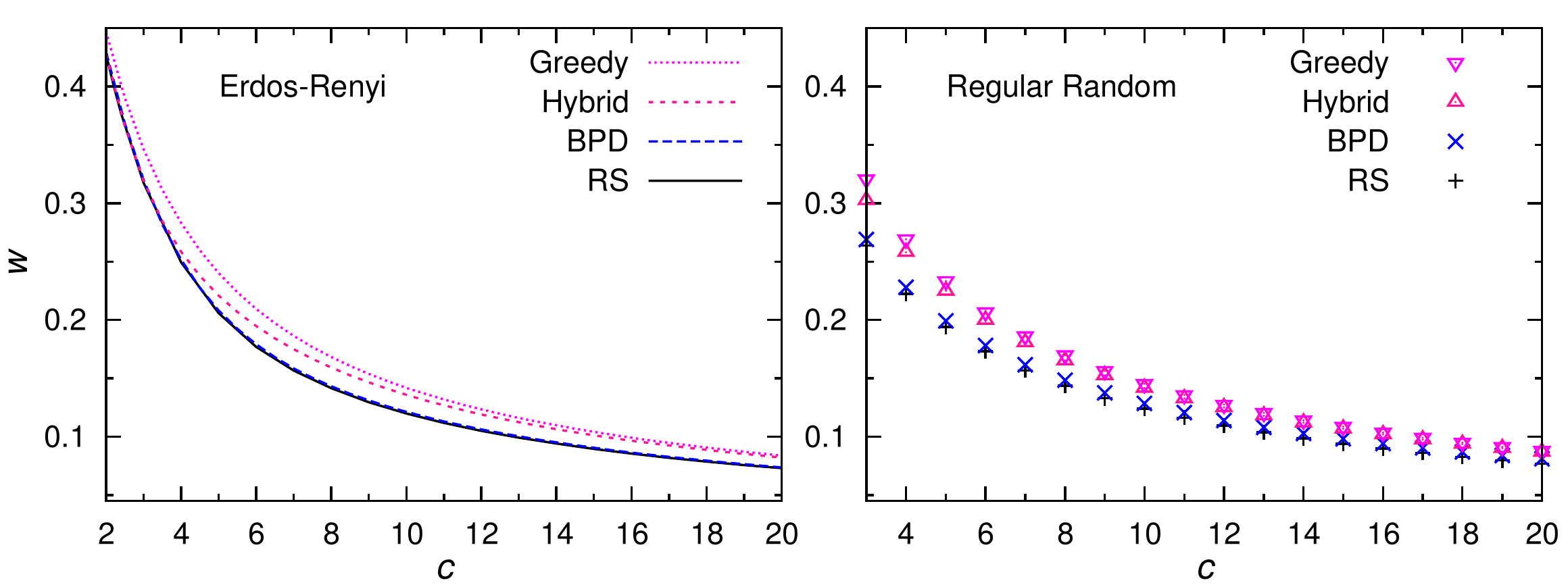}
  \end{center}
  \caption{
    \label{fig:DSminERRR}
    Constructing dominating sets for Erd\"os-R\'enyi networks (left panel)
    and regular random networks (right panel). The relative sizes $w$ of
    dominating sets obtained by a single running of the pure greedy,
    the hybrid, and the BPD algorithm with $x=10$ on $96$ ER or RR 
    network instances of $N=10^5$ and (mean) degree $c$ are compared 
    (fluctuations are of order $10^{-4}$ and are not shown). 
    The ensemble-averaged MDS relative
    sizes obtained by the replica-symmetric mean field theory 
    are also shown (RS).
  }
\end{figure}

\begin{figure}
  \begin{center}
    \includegraphics[angle=270,width=1.0\textwidth]{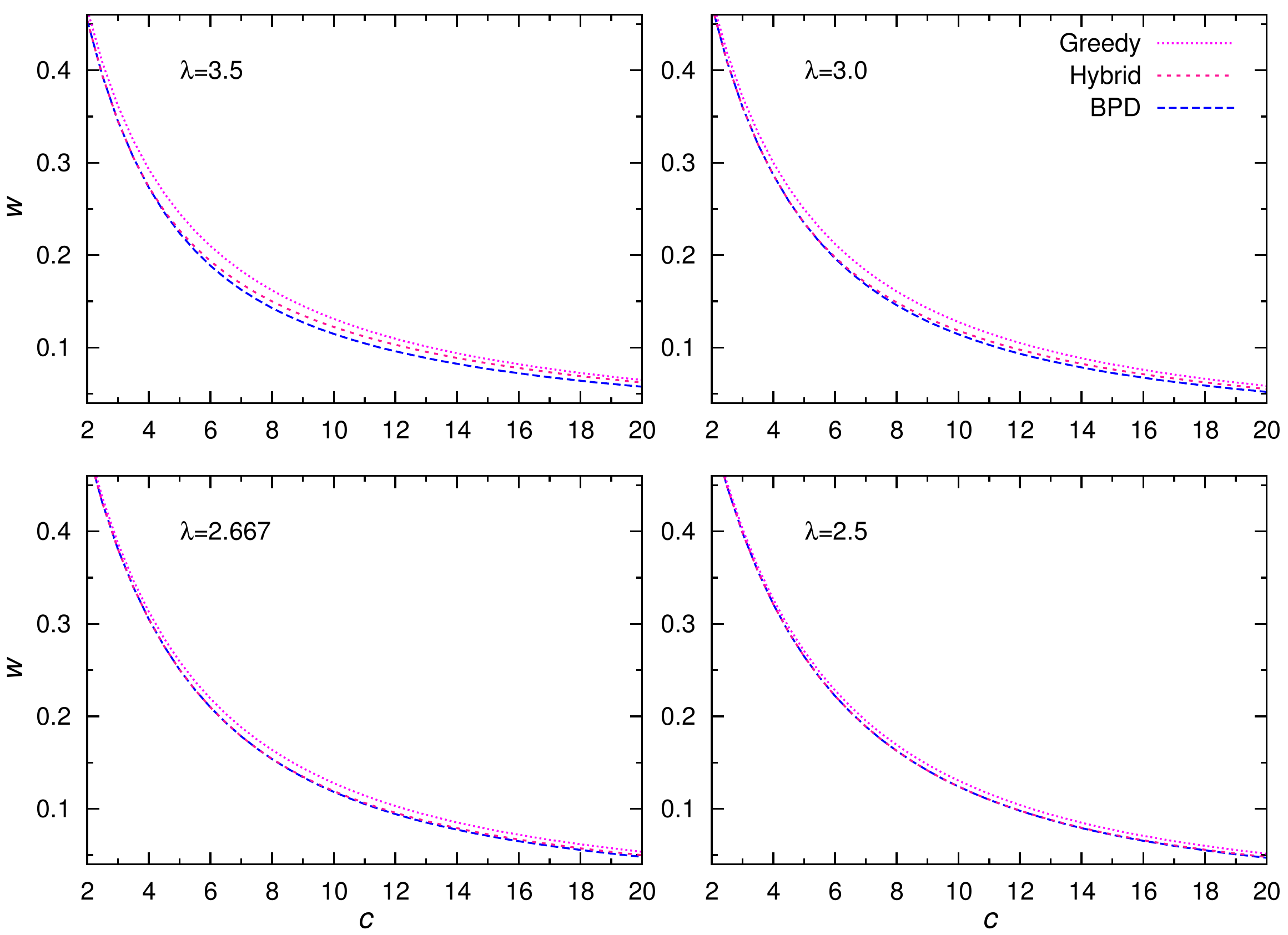}
  \end{center}
  \caption{
    \label{fig:DSminSFstatic}
    Constructing dominating sets for scale-free random networks
    generated through the static model \cite{Goh-Kahng-Kim-2001}.
    The relative sizes $w$ of
    dominating sets obtained by a single running of the pure greedy,
    the hybrid, and the BPD algorithm with $x=10$ on $96$ SF
    network instances of $N=10^5$ and (mean) degree $c$ are compared 
    (fluctuations are of order $10^{-4}$ and are not shown).
    The degree decay exponent is $\lambda =3.5$, $3.0$, $2.667$,
    and $2.5$, respectively.
  }
\end{figure}

Here we introduce an improved local algorithm by combining the GLR process with
the impact-based greedy process. We call this new algorithm the 
{\tt GLR-Impact} hybrid algorithm.
Given an input network $W$ with all the nodes unobserved,
we first carry out the GLR process to simplify $W$ as far as possible.
If all the nodes are observed during this initial GLR process, a MDS of
network $W$ is then constructed.
For the nontrivial case of
some nodes being left unobserved after this GLR process, we first
occupy a randomly chosen node from the subset of highest-impact nodes
and then perform the GLR process again
to further simplify the network as far as possible. We keep repeating such a 
occupying-followed-by-GLR process until there is no unobserved
node left in the network.

The {\tt GLR-Impact} hybrid algorithm is also very easy to implement and
very fast. Its  performance  is demonstrated in 
Fig.~\ref{fig:DSminERRR} for single
ER networks and regular random (RR) networks. (All the nodes of a RR network
have the same integer degree $c$ but the network is otherwise completely
random \cite{Zhao-Zhou-Liu-2013}.)
This hybrid local algorithm outperforms the pure greedy 
algorithm considerably for $c \leq 10$, but it is still inferior to
the belief-propagation-guided decimation (BPD) 
algorithm of section~\ref{sec:bpd}.

\begin{table}
\begin{center}
\begin{footnotesize}
  \begin{tabular}{rrrrrrrr}
    Network & $N$    & $M$ & $d_{max}$ & Core & Greedy & Hybrid & BPD \\ \hline
    RoadEU & $1177$  &  $1417$ & $10$ & $306$ & $428$ & $389$  & $387$ \\
    PPI & $2361$ & $6646$ & $64$ & $17$ & $550$ & $539$ & $539$ \\
    Grid & $4941$ &  $6594$  & $19$ & $603$ & $1564$ & $1485$ & $1481$ \\
    IntNet1 & $6474$ &  $12572$  & $1458$ & $8$ & $660$ & $656$ & $656$ \\
    Author & $23133$  & $93439$ & $279$ & $9052$ &  $3686$ & $3612$ & 
    $3604$ \\
    Citation & $34546$ & $420877$ & $846$ & $11178$ & $3335$ & $3168$ & 
    $3095$ \\
    P2P & $62586$ & $147892$ & $95$ & $35$ & $12710$ & $12582$ & 
    $12582$ \\
    Friend & $196591$ & $950327$ & $14730$ & $6097$ & $42536$ & $41633$ & 
    $41672$ \\
    Email & $265214$ & $364481$ & $7636$ & $470$ & $18183$ & $18181$ & 
    $18181$ \\
    WebPage & $875713$ & $4322051$ & $6332$ & $162439$ & $81288$ & $79928$ &
    $80769$ \\
    RoadTX & $1379917$ & $1921660$ & $12$ & $560582$ & $477729$ & 
    $437503$ & $425774$ \\
    IntNet2 & $1696415$ & $11095298$ & $35455$ & $211244$ & $187592$ &
    $183516$ & $183248$ \\
    \hline
  \end{tabular}
  \end{footnotesize}
\end{center}
\caption{
  \label{tab:Real}
  Results on twelve real-world network instances.
  $N$ and $M$ are, respectively, the total number of nodes and links
  in the network; $d_{max}$ is the maximum node degree of the network;
  the column marked by `Core' records the number of nodes that are left 
  unobserved after the GLR process; the columns marked by
  `Greedy', `Hybrid', and `BPD' record
  the sizes of the dominating sets constructed by a single running of the
  pure greedy, the hybrid, and the BPD algorithm, respectively.
}
\end{table}

We also test the performance of 
the hybrid algorithm on single SF random networks generated through
the static model \cite{Goh-Kahng-Kim-2001} (see
Fig.~\ref{fig:DSminSFstatic}). The {\tt GLR-Impact} algorithm
still outperforms the pure greedy algorithm on these heterogeneous networks,
and its performance approaches that of the BPD algorithm as the network
becomes more and more heterogeneous (i.e., as the decay exponent
$\lambda$ approaches $2.0$ from above).

Real-world networks are often very heterogenous, with a small fraction of 
highly connected nodes \cite{Albert-Barabasi-2002}. 
As a test of the algorithms introduced in this work, 
we apply the GLR process, the pure greedy algorithm, the
hybrid algorithm, and the BPD algorithm to a set of twelve real-world
networks.
Among these network instances, five are infrastructure networks:
European express road network
(RoadEU \cite{Subelj-Bajec-2011}), road network of Texas (RoadTX
\cite{Leskovec-etal-2009}), power grid of western US states (Grid
\cite{Watts-Strogatz-1998}), and two Internet networks at the
autonomous systems level (IntNet1
and IntNet2 \cite{Leskovec-etal-2005});
 three are information
networks: Google webpage network (WebPage \cite{Leskovec-etal-2009}),
European email network (Email \cite{Leskovec-Kleinberg-Faloutsos-2007}),
and research citation network
(Citation \cite{Leskovec-etal-2005});
three are social contact networks:
collaboration network of condensed-matter authors (Author 
\cite{Leskovec-Kleinberg-Faloutsos-2007}), 
peer-to-peer interaction network (P2P
\cite{Ripeanu-etal-2002}),and on-line friendship
network (Friend \cite{Cho-etal-2011});
the remaining one is the biological network
of protein-protein interactions (PPI \cite{Bu-etal-2003}).

The numerical results are summarized in Table~\ref{tab:Real}. 
The GLR process is able to 
simplify these networks considerably. After GLR, the remaining number
of unobserved nodes is often much smaller than the total number $N$ of
nodes in the original network. The BPD algorithm performs slightly
better than the {\tt GLR-Impact} hybrid algorithm, and
both BPD and the hybrid algorithm outperform the pure greedy algorithm 
in all the twelve network instances.

\section{Spin Glass Model and Replica-Symmetric Mean Field Theory}
\label{sec:glass}

If a given network instance $W$ contains an extensive core, the GLR process
can only give a lower bound to the MDS size. 
We now discuss the issue of estimating the MDS size by way of
a mean field theory. We introduce a partition function $Z$ as
\begin{equation}
  \label{eq:model}
  Z= \sum\limits_{\underline{c}}\prod\limits_{i\in W}
  \Bigl\{ e^{-x c_i} \bigl[ 1 - (1-c_i)
    \prod\limits_{j\in\partial i} (1-c_j) \bigr]  \Bigr\} \; ,
\end{equation}
where $\underline{c} \equiv (c_1, c_2, \ldots, c_N)$ denotes one of the
$2^N$ possible occupation configurations, $x>0$ is a re-weighting parameter,
and $\partial i$ denotes node $i$'s
set of adjacent nodes. The constraint of each node $i$ leads to a
multiplication term  $[1-(1-c_i) \prod_{j\in \partial i} (1-c_j)]$, 
which equals to $0$ if $i$ and all its adjacent
nodes are empty and equals to $1$ if otherwise.
The partition function therefore only takes into
account all the dominating sets, and at $x \rightarrow \infty$ it 
is contributed exclusively by the MDS configurations.

\subsection{Replica-Symmetric Mean Field Theory}

We solve the spin glass model (\ref{eq:model}) by a 
RS mean field theory, which can be understood from
the angle of Bethe-Peierls approximation \cite{Mezard-Montanari-2009} or
derived alternatively through partition function expansion 
\cite{Xiao-Zhou-2011,Zhou-Wang-2012}. The marginal probability $q_i^{c}$ of
node $i$'s occupation state being $c$ ($\in \{0,1\}$) is expressed as
\begin{equation}
  \label{eq:qi}
  q_i^{c}  = \frac{ e^{-x c} \prod\limits_{j\in \partial i}
    \sum\limits_{c_j} q_{j\rightarrow i}^{(c_j,c)} - \delta_{0}^{c}
    \prod\limits_{j\in \partial i} q_{j\rightarrow i}^{(0,0)}}
  {\sum\limits_{c_i} e^{-x c_i} 
    \prod\limits_{j\in \partial i} \sum\limits_{c_j} q_{j\rightarrow i}^{(c_j,c_i)}
    -\prod\limits_{j\in \partial i} q_{j\rightarrow i}^{(0,0)} } \; ,
\end{equation}
where the Kronecker symbol $\delta_{m}^{n}=1$ if
$m=n$ and $\delta_{m}^{n}=0$ if otherwise. The quantity 
$q_{j\rightarrow i}^{(c_j,c_i)}$ is defined as the joint probability that node $i$ 
is in occupation state $c_i$ and its adjacent node $j$ is in occupation 
state $c_j$ when the constraint of node $i$ is \emph{not} considered. This
probability can be evaluated through the following
belief-propagation (BP) equation:
\begin{equation}
  q_{j\rightarrow i}^{(c_j, c_i)}  =
  \frac{ e^{-x c_j} \prod\limits_{k\in \partial j\backslash i}
      \sum\limits_{c_k} q_{k\rightarrow j}^{(c_k,c_j)} -
      \delta_{0}^{c_i+c_j} \prod\limits_{k\in \partial j\backslash i} 
      q_{k\rightarrow j}^{(0,0)}}{
    \sum\limits_{c_i^\prime, c_j^\prime} 
    e^{-x c_j^\prime}  \prod\limits_{k\in \partial j\backslash i} 
      \sum\limits_{c_k^\prime} q_{k\rightarrow j}^{(c_k^\prime, c_j^\prime)}
      -\prod\limits_{k\in \partial j\backslash i} q_{k\rightarrow j}^{(0,0)}  } \; ,
  \label{eq:BP}
\end{equation}
where $\partial j\backslash i$ denotes the subset obtained by deleting 
node $i$ from set $\partial j$.

The total free energy $F$ is related to the partition function by 
$F\equiv - (1/x) \ln Z$. According to the RS mean field theory, its
expression is
\begin{equation}
  \label{eq:F0}
  F = \sum\limits_{i \in W} f_{i} - \sum\limits_{(i,j) \in W} f_{(i,j)} \; ,
\end{equation}
where $f_i$ and $f_{(i,j)}$ are the free energy contributions of a node $i$ 
and a link $(i,j)$ between nodes $i$ and $j$:
\begin{eqnarray}
 & & \hspace*{-0.9cm} f_i = -\frac{1}{x} \ln \Bigl[
  \sum\limits_{c_i} e^{-x c_i} \prod\limits_{j\in \partial i} 
  \sum\limits_{c_j} q_{j\rightarrow i}^{(c_j,c_i)}
  - \prod\limits_{j\in \partial i} q_{j\rightarrow i}^{(0,0)} \Bigr] \; ,
  \label{eq:fi}\\
 & & \hspace*{-0.9cm} f_{(i,j)}  =  -\frac{1}{x} \ln \Bigl[
    \sum\limits_{c_i, c_j} q_{i\rightarrow j}^{(c_i, c_j)}
    q_{j\rightarrow i}^{(c_j, c_i)} \Bigr] \; .
\label{eq:fij}
\end{eqnarray}
From Eqs.~(\ref{eq:F0}) and (\ref{eq:qi}) we can compute the free energy
density $f \equiv F/N$ and the mean occupation fraction 
$w = (1/N)\sum_{i\in W} q_i^{+1}$. The entropy density of the system is then 
evaluated as $s =  (w - f) x$.

\subsection{Belief-Propagation Iterations}

According to Eq.~(\ref{eq:BP}) each probability distribution 
$q_{j\rightarrow i}^{(c_j, c_i)}$ has the property that 
$q_{j\rightarrow i}^{(1,1)} = q_{j\rightarrow i}^{(1,0)}$. 
Therefore in the numerical computations 
$q_{j\rightarrow i}^{(c_j, c_i)}$ can be represented by three non-negative real 
numbers $q_{j\rightarrow i}^{(0,0)}$, $q_{j\rightarrow i}^{(0,1)}$, and 
$q_{j\rightarrow i}^{(1,0)}$, which satisfy in addition the normalization
condition
\begin{equation}
  \label{eq:unity}
  q_{j\rightarrow i}^{(0,0)} + 
  q_{j\rightarrow i}^{(0,1)} +
  2 q_{j\rightarrow i}^{(1,0)} = 1 \; .
\end{equation}

\begin{figure}
\begin{center}
   \includegraphics[angle=270,width=1.0\textwidth]{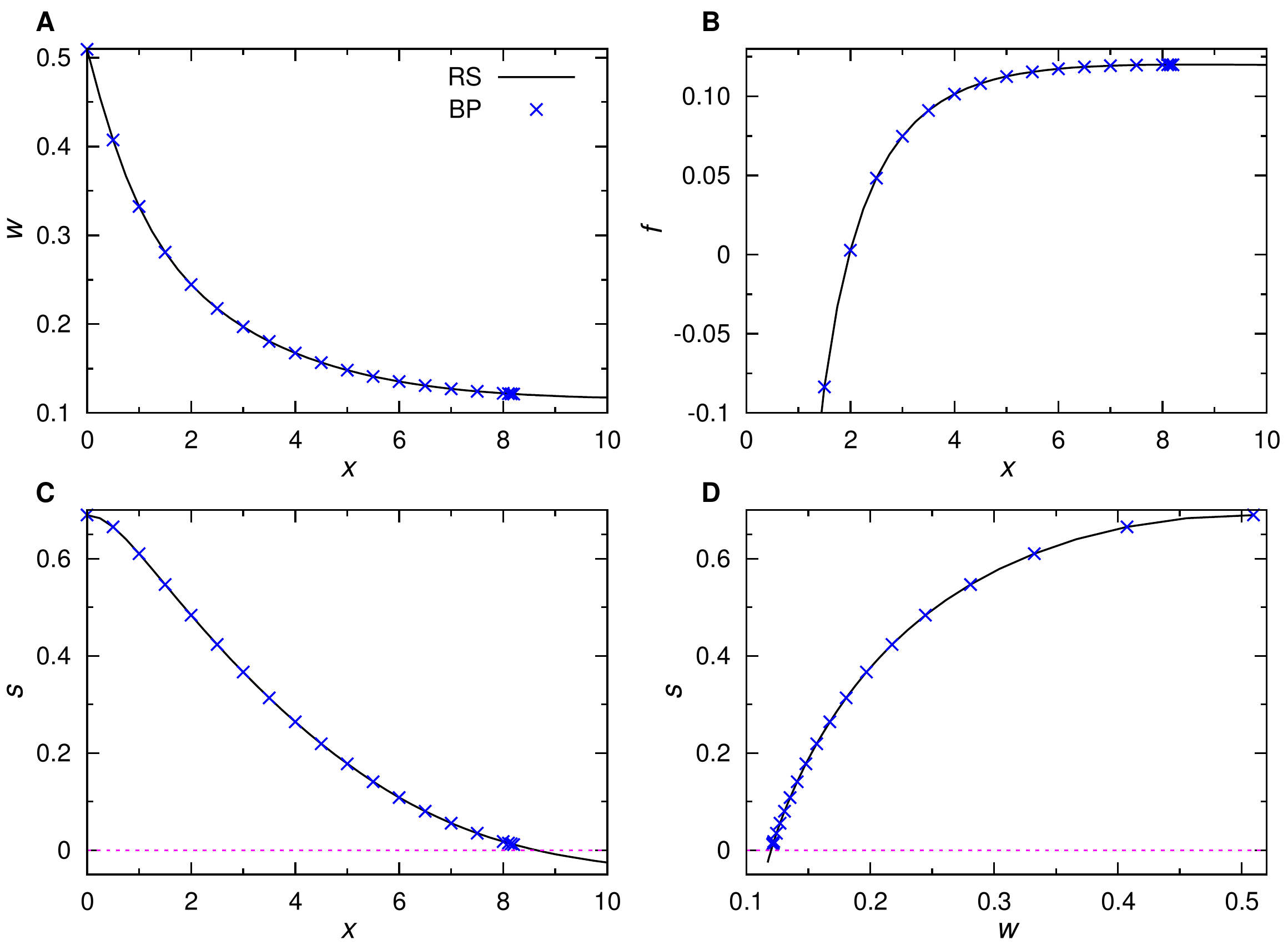}
\end{center}
\caption{\label{fig:RSforER10}
  Replica-symmetric (RS) mean field theory and belief-propagation (BP) results
  for ER random networks of mean degree $c=10$. The RS results are obtained by
  population dynamics simulations, while the BP results are obtained on a
  single ER network instance of $N=10^6$ nodes. The BP iteration converges
  to a fixed point only for $x<8.22$. (A) occupation fraction $w$; (B) free
  energy density $f$; (C) entropy density $s$; (D) entropy density $s$ as a
  function of occupation fraction $w$.
}
\end{figure}

We initialize $q_{j\rightarrow i}^{(c_j, c_i)}$ and $q_{i\rightarrow j}^{(c_i,c_j)}$ 
for each link $(i,j)$ of the network between two nodes $i$ and $j$, for
example setting  
$q_{j\rightarrow i}^{(0,0)}=q_{j\rightarrow i}^{(0,1)}=q_{j\rightarrow i}^{(1,0)}=1/4$. 
We then perform BP iteration a number $T$ of times at a given value of 
the re-weighting parameter $x$,
until a fixed-point solution of Eq.~(\ref{eq:BP}) is reached or $T$ exceeds
a pre-specified number (e.g., $1000$).
In each BP iteration step we treat all the nodes of the network in a
random order. When node $j$ is examined, the output messages 
$q_{j\rightarrow i}^{(c_j,c_i)}$ to all its adjacent nodes $i\in \partial j$ are
updated according to Eq.~(\ref{eq:BP}). 
The difference  $\Delta_{j\rightarrow i}(t)$
between an updated message $q_{j\rightarrow i}(t)$ at the $t$-th BP step and 
the old message $q_{j\rightarrow i}(t-1)$ at the
$(t-1)$-th BP step is defined as
\begin{eqnarray}
  \Delta_{j\rightarrow i}(t) & \equiv &  
  |q_{j\rightarrow i}^{(0,0)}(t)-q_{j\rightarrow i}^{(0,0)}(t-1)| +
|q_{j\rightarrow i}^{(0,1)}(t)-q_{j\rightarrow i}^{(0,1)}(t-1)| \nonumber \\
& & \quad  + 
2  |q_{j\rightarrow i}^{(1,0)}(t)-q_{j\rightarrow i}^{(1,0)}(t-1)| \; .
\end{eqnarray}
If the maximal value among the set of $2 M$ difference values
$\{\Delta_{j\rightarrow i}(t)\}$ is less than certain pre-specified threshold 
value (e.g., $10^{-3}$ or even smaller), then BP iteration is regarded as 
being converged. At a fixed point of Eq.~(\ref{eq:BP})
we then compute the free energy density $f$,
the mean occupation fraction $w$, and the entropy
density $s$ through the RS mean field theory. As an example, we show
in Fig.~\ref{fig:RSforER10}  the results obtained on a single 
ER random  network of size $N=10^6$ and mean degree $c=10$.

For ER networks with mean degree $c>2.41$ and regular random 
networks with integer degree $c\geq 3$, we find that when the
re-weighting parameter $x$ is  larger than certain threshold value,
BP iteration is unable to converge to a 
fixed point. Such a non-convergence phenomenon indicates that, when the
random network system has an extensive core, it will be in a spin glass
phase at sufficiently large values of $x$. 
Systematic theoretical investigations on this spin glass phase will be
reported in another publication.

\subsection{Ensemble-Averaged Properties}

A random network ensemble is characterized by a degree distribution 
$P(d)$. We perform population dynamics simulations using
Eqs.~(\ref{eq:BP}), (\ref{eq:qi}) and (\ref{eq:F0}) to
obtain ensemble-averaged results.
First, we create a long array $A$ of $\mathcal{N}$ (e.g., $10^5$)
elements to store a set of
messages, each of which represents a probability distribution 
$q_{j\rightarrow i}^{(c_j,c_i)}$ in the form of a three-dimensional vector 
satisfying Eq.~(\ref{eq:unity}): 
$q_{j\rightarrow i}\equiv 
(q_{j\rightarrow i}^{(0,0)}, q_{j\rightarrow i}^{(0,1)},
q_{j\rightarrow i}^{(1,0)})$. We then repeatedly update elements of this array
by the following procedure:
(1) generate a random integer $d\geq 1$ according to the degree probability
distribution $Q(d)$; (2) draw $(d-1)$ elements $q_{k\rightarrow j}$
from array $A$ uniformly at random, and then use these $(d-1)$ elements 
as input messages to Eq.~(\ref{eq:BP}) to compute a new message 
$q_{j\rightarrow i}$;
(3) replace a randomly chosen element of array $A$ with this new message.
The message array $A$ is expected to reach a steady state after it is 
updated a sufficient number of times (e.g., after each element of this 
array is updated $10,000$ times on average).

We then keep updating the message array $A$ and at the same time compute
the thermodynamic quantities $f$, $w$, and $s$. For example, the free
energy density $f$ is obtained by
\begin{equation}
  f=\overline{f_i} - \frac{c}{2} \overline{f_{(i,j)}} \; ,
\end{equation}
where $\overline{f_i}$ is the average of the free energy node contribution
$f_i$ over all the nodes, and $\overline{f_{(i,j)}}$ is the average of the
free energy link contribution $f_{(i,j)}$ over all the links. We generate many
samples of $f_i$ and $f_{(i,j)}$ to compute their averages
$\overline{f_i}$ and $\overline{f_{(i,j)}}$. 
The procedure of obtaining a sample of  $f_i$ is the same
as that of updating an element of the message $A$,
the only difference being that the degree $d_i$
of node $i$ should be generated according to the distribution $P(d)$ instead
of $Q(d)$. A sample of $f_{(i,j)}$ is obtained very easily through
Eq.~(\ref{eq:fij}) by picking two messages $q_{j\rightarrow i}$ 
and $q_{i\rightarrow j}$ uniformly at random from the message 
array $A$.

For ER random networks with mean degree $c=10$, we compare in
Fig.~\ref{fig:RSforER10} the results obtained by this 
RS population  dynamics with the results obtained by BP iteration 
on  a single network instance. The ensemble-averaged results are 
in perfect agreement with the BP iteration
results (provided the BP iteration is able to converge).

The entropy density $s$ as a function of the mean
occupation fraction $w$ can be obtained from these RS population dynamics
results (see for example Fig.~\ref{fig:RSforER10}D). In some random 
network systems, the entropy density
$s$ become negative if $w$ decreases below certain threshold value 
$w_0$, indicating that there is no dominating set with relative
size below $w_0$.  We therefore take the value $w_0$ 
as the ensemble-averaged MDS relative size. For 
ER networks of $c=10$, we obtain from Fig.~\ref{fig:RSforER10} that
$w_0 \approx 0.120$ (the corresponding value of $x$ is
$x\approx 8.637$). In some other random network systems (e.g., ER
random networks with $c<2.41$, before the core percolation transition),
the entropy density $s$ approaches a non-negative limiting value as $w$
approaches a limiting value $w_0$ from above. For these latter cases, we
simply take $w_0$ as the ensemble-averaged MDS relative size.

The ensemble-averaged results on the MDS sizes of ER and RR networks
are shown in Fig.~\ref{fig:DSminERRR}. For ER networks with mean 
node degree $c<2.41$ (before the core-percolation transition),
the RS mean field results coincide with the results predicted by the
core percolation theory. When the random network contains an extensive
core, the results obtained by the pure greedy algorithm and the
{\tt GLR-Impact} algorithm are higher than the RS mean field predictions, 
but the results obtained by the BPD algorithm of the next section are 
very close
to the RS mean field predictions.

\section{Belief-Propagation-Guided Decimation Algorithm}
\label{sec:bpd}

For a given network $W$, the RS mean field theory gives an estimate for the
occupation probability $q_{i}^{+1}$ of each node $i$, see Eq.~(\ref{eq:qi}).
Such information is exploited in a BPD algorithm to construct a
near-optimal dominating set. (Such an algorithm and its extensions have
already been successfully applied to many other combinatorial optimization
problems, e.g., the $K$-satisfiability problem
\cite{Mezard-etal-2002,Krzakala-etal-PNAS-2007} and
the vertex-cover problem \cite{Zhao-Zhou-2014}.)
At each round of the BPD process, unoccupied
nodes with the highest estimated occupation probabilities are added to
the dominating set, and the occupation probabilities for the remaining
unoccupied nodes are then updated.

If a node $j$ is unobserved (it is empty and has no adjacent occupied node), 
the output message 
$q_{j\rightarrow i}^{(c_j, c_i)}$ on the link $(j,i)$ between $j$ and
node $i$
is updated according to Eq.~(\ref{eq:BP}). On the other hand, if node $j$ 
is empty
but observed (it has at least one adjacent occupied node), this node then 
presents no restriction to the occupation states of all its unoccupied
neighbors. For such a node $j$, the output message 
$q_{j\rightarrow i}^{(c_j,c_i)}$  on the link $(j,i)$ is then updated according
to the following equation:
\begin{equation}
  \label{eq:qji2}
  q_{j\rightarrow i}^{(c_j,c_i)} = 
  \frac{e^{-x c_j} \prod\limits_{k\in \partial j\backslash i} \sum\limits_{c_k} 
    q_{k\rightarrow j}^{(c_k,c_j)}}{\sum\limits_{c_j^\prime, c_i^\prime} 
    e^{-x c_j^\prime} \prod\limits_{k\in \partial j\backslash i} \sum\limits_{c_k^\prime}
    q_{k\rightarrow j}^{(c_k^\prime,c_j^\prime)}} \; .
\end{equation}
Similar to Eq.~(\ref{eq:qji2}), the marginal probability 
distribution $q_i^{c_i}$ for an
observed empty node $i$ is evaluated according to
\begin{equation}
\label{eq:qiobs}
  q_{i}^{c_i} = 
  \frac{ e^{-x c_i}\prod\limits_{j\in \partial i} \sum\limits_{c_j} 
    q_{j\rightarrow i}^{(c_j,c_i)}}{\sum\limits_{c_i^\prime} e^{-x c_i^\prime}
    \prod\limits_{j\in \partial i} \sum\limits_{c_j^\prime}
    q_{j\rightarrow i}^{(c_j^\prime,c_i^\prime)}} \; .
\end{equation}
It is easy to verify from Eq.~(\ref{eq:qji2}) that 
$q_{j\rightarrow i}^{(0,0)}=q_{j\rightarrow i}^{(0,1)}$
and $q_{j\rightarrow i}^{(1,0)}=q_{j\rightarrow i}^{(1,1)}$. Notice that if
all the nodes in the set $\partial j\backslash i$ are observed, then
we derive from Eq.~(\ref{eq:qji2}) that $q_{j\rightarrow i}^{(0,0)}=
q_{j\rightarrow i}^{(1,0)}=q_{j\rightarrow i}^{(0,1)}=q_{j\rightarrow i}^{(1,1)}=
1/4$. Because of this property, we need only to consider the links between
unobserved nodes and the links between unobserved and observed
nodes. All the other links (which are between observed nodes) do not 
need to be considered in the BP iteration equations (\ref{eq:BP}) 
and (\ref{eq:qji2}).

We implement the BPD algorithm as follows:
\begin{enumerate}
\item[(0)]
  Input the network $W$, set all the nodes to be empty and unobserved
  and set all the
  probability distributions $q_{j\rightarrow i}^{(c_j,c_i)}$ to be the 
  uniform distribution.
  Set the re-weighting parameter $x$ to a sufficiently large value (e.g.,
  $x=10$). Then perform the BP iteration a number $T_0$ of rounds (e.g.,
  $T_0=200$). 
  After these $T_0$ iterations we compute the occupation probability
  $q_i^{+1}$ of each node $i$ using Eq.~(\ref{eq:qi}).
  
  \item[(1)] 
    Then occupy a small fraction $r$ (e.g., $r=0.01$) of the unoccupied
    nodes that having the highest estimated occupation probabilities.
    
  \item[(2)]
    Then simplify network $W$ by first deleting all the links 
    between observed nodes, and then deleting all the 
    isolated observed nodes.
    
  \item[(3)]
    If the resulting network $W$ still contains unobserved nodes,
    we perform BP iteration for a number of $T_1$ rounds (e.g., $T_1=10$).
    The output message of an node $i$ is updated either according to
    Eq.~(\ref{eq:qi}) or 
    according to Eq.~(\ref{eq:qiobs}), depending on whether $i$ is
    unobserved or observed.
    We then repeat operations (1)--(3)  until all
    the nodes are observed.
\end{enumerate}
In addition, we may first carry out the GLR process
to simplify the network $W$ as
far as possible before running the BPD process. For real-world networks
with some nodes being highly connected, we find that such a GLR simplifying
step reduces the BPD running time considerably and also slightly 
reduces the size of the constructed dominating set.

The results of the BPD algorithm for random networks and for
real-world networks are compared with the results obtained by the
local heuristic algorithms in
Fig.~\ref{fig:DSminERRR}, Fig.~\ref{fig:DSminSFstatic}, and
Table~\ref{tab:Real}. For ER and RR random networks,
the BPD algorithm considerably beats both the
pure greedy algorithm and the {\tt GLR-Impact} hybrid algorithm; for
very heterogeneous (e.g., scale-free) networks,
the BPD algorithm only slightly outperforms the {\tt GLR-Impact} algorithm.

\section{Discussions}
\label{sec:conclude}

In this work, we proposed two heuristic algorithms 
(a {\tt GLR-Impact} local algorithm and
a BPD message-passing algorithm) and presented a 
core percolation theory and a 
replica-symmetric mean field theory for solving
the network dominating set problem algorithmically and theoretically. 
We found that the GLR process may lead to a core percolation transition
in the network (see Fig.~\ref{fig:erglr} and Fig.~\ref{fig:sfstaticglr}).
Our numerical results shown in Fig.~\ref{fig:DSminERRR}, 
Fig.~\ref{fig:DSminSFstatic} and Table~\ref{tab:Real} suggested that
the {\tt GLR-Impact} algorithm and the BPD algorithm can construct
near-optimal dominating sets for random networks and real-world networks.

There are many theoretical issues remaining to be investigated.
An easy extension of the core percolation theory is to consider 
GLR with a subset of initially occupied nodes. By optimizing this initial
subset (e.g., following the methods of \cite{Altarelli-Braunstein-DallAsta-Zecchina-2013b,Altarelli-Braunstein-DallAsta-Zecchina-2013,Guggiola-Semerjian-2014}),
we may reach an improved lower-bound to the MDS size. 
Core percolation on degree-correlated random
networks \cite{Hasegawa-Takaguchi-Masuda-2013} and in the more general 
lattice glass problem \cite{Biroli-Mezard-2002} are also very interesting.
When the random network has an extensive core, we observed that
the belief-propagation equation (\ref{eq:BP}) fails to converge at 
large values of the re-weighting parameter $x$ (see Fig.~\ref{fig:RSforER10}),
indicating a spin glass phase transition.
A systematic study of the spin glass phase will be carried out using the
first-step replica-symmetry-breaking mean field theory 
\cite{Mezard-Parisi-2001,Mezard-Montanari-2006,Krzakala-etal-PNAS-2007}, 
which may in addition
offer an improved estimate on the ensemble-averaged MDS size.
The possible deep connections between core percolation and the
complexity of the random MDS problem  will also be addressed by adapting 
the long-range frustration theory \cite{Zhou-2005a,Zhou-2012}.

The methods of this work can be readily extended to the MDS problem of
directed networks. Our theoretical and algorithmic results on the directed
MDS problem will soon be reported in an accompanying paper
\cite{Habibulla-Zhao-Zhou-2015}.
A more challenging problem is the connected dominating set 
problem \cite{Du-Wan-2013} which has the additional constraint that
the nodes in the dominating set should induce a connected subnetwork.
Our present work may stimulate further theoretical studies on this hard
problem.

\section*{Acknowledgments}

Part of this work was done when H.-J.Zhou was participating in 
the ``Collective Dynamics in Information Systems 2014'' Program of 
the Kavli Institute for Theoretical Physics China (KITPC).
H.-J. Zhou thanks Chuang Wang for a helpful discussion, and 
Alfredo Braunstein, Yang-Yu Liu, Federico Ricci-Tersenghi, and
Yi-Fan Sun for helpful comments on the manuscript;
J.-H. Zhao and H.-J. Zhou thank Prof. Zhong-Can Ou-Yang for support.

\end{document}